# A generalized network level disruption strategy selection model for urban public transport systems

**Qi Liu, Joseph Y. J. Chow**[*]

*Department of Civil and Urban Engineering, New York University, Brooklyn, NY, 11201, USA*
[*]Corresponding author email: joseph.chow@nyu.edu

**Abstract**
A fast recovery from disruptions is of vital importance for the reliability of transit systems. This study presents a new attempt to tackle the transit disruption mitigation problem in a comprehensive and hierarchical way. A network level strategy selection optimization model is formulated as a joint routing and resource allocation (nJRRA) problem. By constraining the problem further into an $\epsilon$-constrained nJRRA problem, classic solution algorithms can be applied to solve the quadratically constrained quadratic program (QCQP). On top of this "basic model", we propose adding a decision to delay the resource allocation decisions up to a maximum initiation time when the incident duration is stochastic. To test the models, a quasi-dynamic evaluation program with a given incident duration distribution is constructed using discretized time steps and discrete distributions. Five different demand patterns and four different disruption duration distributions (20 combinations) are tested on a toy transit network. The results show that the two models outperform benchmark strategies such as using only line level adjustment or only bus bridging. They also highlight conditions when delaying the decision is preferred.

# 1 Introduction

## 1.1 Background

Daily transit operations encounter various types of disruptions, like track failure, rolling stock failure, intrusions, medical emergencies, weather/nature disasters, etc. Serious service degeneration may propagate through the network and last for hours. Given the importance of transit service reliability, the application of recovery models and algorithms for real-time disturbance and disruption management is considered a key element for improving the service and reliability of transit systems (Cacchiani et al., 2014). This is true for urban public transport in general. There are many strategies in use today for a typical transit system. However, it is not always obvious how to find the optimal combination of strategies in real-time.

## 1.2 Motivations

Typical real-time transit management system found in Ceder (2016), Ben-Akiva et al. (2001), Dessouky et al. (2003), and Cats (2011) have similar structures. Collected real-time data are sent to a state estimation/prediction model. The output (current states estimations, predictions) of this model is redirected to a strategy selection model, where the optimal action is sought. Within the system, a real-time transit disruption mitigation strategy or policy is determined in several different ways: by optimization, by looking up a contingency table, or just by experience from expert knowledge. It involves many different roles like scheduler, dispatcher, driver, infrastructure maintainer, etc. The strategy is transmitted to all relevant parties for execution.

The exact set of feasible disruption mitigation strategies may differ from system to system or even from line to line because of the availability of crossings, parallel tracks, backup vehicles and staff, user acceptance, etc. Ceder (2016) gives a comprehensive list of real-time control strategies:

- Holding the vehicle (at terminal or at mid-route point);
- Skip-stop operation;
- Adding a reserve vehicle;
- Changes in speed (not above the lawful speed limit);
- Interlining operation;
- Deadhead operation;
- Short-turn operation;
- Short-cut operation;
- Leapfrogging operation with the vehicle ahead.

Other strategies include bus bridging for metro (Kepaptsoglou and Karlaftis, 2009), emergency lines (Cadarso et al., 2013), service network redesign (Kiefer et al., 2016), and cancellation/addition of tasks (Thengvall et al., 2000).

Given all these mitigation strategies, some represent minor fine-tunings of the current service, like holding or skip-stop decisions. These decisions can be made locally and are relatively easy to implement. Others, like inter-lining and bus-bridging, call for wider collaborations and demand more efforts. Transit agencies tend to avoid them unless the situation is serious enough. There is a need for a model to optimize the strategy selection step shown in Figure 2. For the purposes of this study, we define transit *disruption* as an unexpected event that requires an operator or user to adjust their original schedules. Some disruptions only have minor effects, such as a bus being delayed for a few minutes. Others call for substantial changes of the original schedule, say a tunnel shutdown. We classify disruptions by the following definitions, which are adapted from the study of Clausen et al. (2010) from the airline industry. We consider minor and major disruptions. The latter is defined to be sufficiently significant to trigger costly strategies and this is when our proposed strategy selection model would apply. Determination of the threshold between these disruptions falls on the local agency as it may vary from agency to agency. This study, like all the studies mentioned in the literature, focuses on major disruptions that nonetheless allow the system to remain operational. We are not studying a disaster evacuation problem that is targeting a transit system in which the focus is on safely evacuating passengers out of the system (Yazdani, 2020).

**Definition** (major disruption): A *major disruption* of an urban transit system is an event or a series of events that renders the planned schedules for a predefined threshold of users ($N_u$), service tasks ($N_t$), vehicles ($N_v$), or crews ($N_c$) infeasible. To be exact, a major disruption event $E$ is defined as:
$$E = E_u \cup E_t \cup E_v \cup E_c$$
where
$E_u$ := the event that the number of users whose schedules become infeasible is no less than $N_u$;
$E_t$ := the event that the number of tasks to be added, deleted, or with schedules being deviated no less than $\epsilon_t$, is no less than $N_t$;
$E_v$ := the event that the number of vehicles whose routes must deviate from their original routes is no less than $N_v$;
$E_c$ := the event that the number of crews whose assignments to tasks are changed is no less than $N_c$.

**Definition** (minor disruption): A *minor disruption* is an event or a series of events that is not a major disruption.

We propose a new strategy selection model to tackle major disruptions by selecting strategies at the network level to allocate resources. Strategies like bus bridging, inter-lining, short-turning, service line redesigning, service run adjustment, are all considered by our model. This is done by mapping the strategies into equivalent fleet allocation decisions.

### *1.3 Related studies in disruption mitigation strategies*
There are many studies on disruption mitigation strategies. Reviews of disruption management for passenger railway transportation can be found in Jespersen-Groth et al. (2009) and Cacchiani et al. (2014). This study is not concerned with planning level strategies like Jin (2014), Mudigonda et al. (2019) and Zeng et al. (2021).

*Minor service adjustments problem*
Newell (1971), Wirasinghe (2002) studied headway control using queueing theory. Osuna and Newell (1972) presented control strategies for holding a vehicle for a hypothetical bus-route loop with only one service and control point. Barnett (1974) discussed vehicle holding strategies at control points to deal with randomness. Turnquist (1982) and Furth (1995) proposed analytic models to adjust the headways after a disruption. Hickman (2001) described an analytic model that determines the optimal vehicle holding time at a control stop along a transit route. Cats et al. (2011) and Berrebi et al. (2018) compared bus holding control strategies, like a schedule-based holding strategy, minimum headway requirement, and even-headway strategy, all evaluated using simulation. Adamski and Turnau (1998) addressed the problem of minimizing schedule deviations on a route. O'Dell and Wilson (1999) presented formulations for disruption control problems with holding and short-turning strategies for systems with more than one branch.

Joint optimization models involving multiple strategies like holding, stop-skipping, expressing, short-turning, and deadheading, are usually formulated as mixed integer programming problems. Li et al. (1992) optimized departure time and skip-stop decisions to minimize the waiting time along a horizon. Turnquist (1982) controlled vehicle speed to recover to the original schedule. Eberlein et al. (1999) proposed a deterministic optimization model that includes control strategies like deadheading, expressing, and holding. Fay and Schnieder (1999) applied fuzzy Petri nets to formulate a knowledge-based decision support system for transit tactical-level decisions like holding. Shen and Wilson (2001) described an integrated real-time disruption control model formulated as a mixed integer nonlinear program for rail transit systems. It included holding, expressing and short-turning strategies. Su et al. (2020) proposed a metro re-scheduling model based on Q-learning, a type of reinforcement learning technique. Gao et al. (2016) proposed a mixed integer optimization model to find the optimal departure time and skipping-stop strategies after disruptions for a metro system. Wang et al. (2015) proposed a nonlinear model to find the departure time and splitting rates solved by evolutionary algorithms. Berger et al. (2011) studied whether a train should wait for a delayed incoming train to facilitate transfer. The problem is represented using an event-graph and formulated as a variant of the uncapacitated multi-commodity flow problem. The major objective of the model is the satisfaction of network passengers. Sáez et al. (2012) proposed an optimization model for real-time bus holding and expressing control, solved by genetic algorithm. Hassannayebi et al. (2021) proposed an event based simulation and used neighborhood search to optimize short-turning and stop-skipping decisions. Zhu et al. (2022) proposed a mixed-integer nonlinear robust optimization program to find the short-turning and train circulation decisions. Farrag et al. (2021) focused on microscopic vehicle motion under disruption; V2X technology is applied to help vehicles to pass through the road incident bottleneck more smoothly.

*Service run adjustment problem*
Run addition or removal changes the headways resulting in larger consequences than holding strategies. If a run gets canceled, the current vehicle or crew plan may become infeasible. Most airline disruption mitigation models jointly consider run cancellation and delay options. Cost or profit are associated with each potential run. The routes of aircrafts are optimized to maximize the total profits. The model is formulated as an integer linear program (ILP) by Jarrah et al. (1993), Thengvall et al. (2000). The latter proposed an integer linear programming model using a time-space network. Zhan et al. (2015) studied the rescheduling of railway traffic on a high-speed railway line in case of a complete blockage. A mixed integer program was proposed and solved by two-stage optimization approach. Veelenturf et al. (2017) proposed a model for the joint rescheduling of timetable and rolling stock for a railway system, solved by heuristic algorithm. Yuan et al. (2022) proposed a model to jointly optimize the assignment of users and transit schedules. The problem is formulated as a MILP and solved by CPLEX. Yuan et al. (2023) proposed integrated optimization approach for passenger flow control and metro scheduling considering skip-stop patterns. Passenger flow control measures include closing a part of the automatic fare gates, setting railings, closing entrances and exits, closing transfer channels, etc. The model is formulated as a mixed integer program.

*Service line redesign problem*
There are only a few studies on real-time service line redesign. Kiefer et al. (2016) proposed a mixed integer programming model to respond to serious disruptions by redesigning the lines in a particular region around the disruption and adjusting the frequencies. In Cadarso et al. (2013), lines can be canceled and emergency lines added. The rolling stock is jointly optimized.

*Substitution service design problem*
Substituting a service by another mode may occur when a disruption disables the service locally. The bus is the most popular choice for substituting other modes (i.e. bus bridging). The bus bridging problem is similar to the transit route network design problem (TRNDP). It typically consists of three steps: first, a heuristic method is used to generate candidate routes; then an optimization model is employed to find their frequencies; and lastly, the routes for individual buses are optimized (Kepaptsoglou and Karlaftis, 2009, Gu et al., 2018, Jin et al., 2016, Kang et al., 2019). Bus bridging routes are generated using a shortest path algorithm and subsequently modified through a heuristic approach. Gu et al. (2018) developed a two-stage integer linear programming model to flexibly allocate and schedule buses to a set of shuttle bus routes during unexpected metro disruptions. Zhang and Lo (2018) investigated the optimal initiation time for substitute bus service. Cheng and An (2021) studied integrated optimization of bus bridging routes and train timetables under rail disruption.

*Vehicle/crew rescheduling problem*
Recovering from serious disruptions may require changes to the timetable, the rolling stock, as well as the crew duties. The vehicle and crew rescheduling problems are very similar. They are both about finding paths to cover a set of tasks. They are usually formulated as multi-commodity minimum cost flow problems (Desrosiers et al., 1995, Ribeiro and Soumis, 1994, Löbel, 1997, Mesquita and Paixão, 1999, Huisman et al., 2004). Alternatively, they can be formulated as set partition/covering problems if trajectories are enumerated (Friberg and Haase, 1999, Mingozzi et al., 1999, Yu et al., 2003, Mesquita and Paias, 2008). Visentini et al. (2014) reviewed the vehicle rescheduling problem for road traffic, railway, and airlines. The set of possible routes of a realistic network is too large to enumerate. Hence, column generation is often used to solve the vehicle/crew rescheduling problem (Yu et al., 2003, Stojković et al., 1998, Nissen and Haase, 2006, Medard and Sawhney, 2007, Lettovský et al., 2000). Li et al. (2007) have done a series of studies on the vehicle rescheduling problem (VRSP). It is based on the Single Depot Vehicle Scheduling Problem (SDVSP), which assigns vehicles to a set of predetermined trips with fixed starting and ending times with an objective of minimizing capital and operating costs. Li et al. (2009) proposed a single depot vehicle rescheduling model solved by a Lagrangian relaxation-based heuristic.

Lai and Leung (2018) proposed a joint line frequency, vehicle, and crew schedule optimization model under a rolling horizon framework. The objective is to maximize the route frequency and to minimize the crew overtime and meal-break delay. The disruptions include unexpected traffic conditions, vehicle breakdown, staff leave, etc. Carosi et al. (2015) and Malucelli and Tresoldi (2019) proposed joint vehicle and crew optimization models for disruption management. When an irregularity is detected, a simulation-based optimization tool is used to select from the set of actions which includes holding, short-turning, expressing, and speed adjustment.

For railway systems, there are extra blocking rules to ensure safety. The problem of finding the best way of arranging a set of operations to minimize the span of execution is a job shop scheduling problem with blocking constraints. Mascis and Pacciarelli (2002) proposed a branch and bound (B&B) algorithm with the technique of immediate selection (or dynamic selection) to decrease the branching factor. D'ariano et al. (2007) extended the B&B algorithm by adding a preprocessing phase to compute the static implication sets for each arc; these sets help to reduce the number of branches. In the study of Walker et al. (2005), the task sequence of each vehicle is fixed; the decision variables are departure times, holding times at stations, the track crossing precedence variables, and the crew's path variables.

*1.4 Research gaps and contributions*

Firstly, disruption durations are typically unknown in advance. There is a trade-off between user cost and operator cost. User demands are usually stochastic and only partially observable. Disruption mitigation considering all these stochastic factors has not been fully investigated.

Secondly, most previous studies on urban transport are line level models to limit the size of the problem. However, passengers re-route on the whole network, and resources (like crews and vehicles) are distributed across the network. Models for intercity trains or airlines are indeed network level, but the disruption mitigation strategies for these systems are not as rich as urban public transport. For example, urban public transport needs more precise controls when it comes to dwell times and headways; urban public transport has bus-bridging options, etc.

There is a need for an efficient disruption strategy selection model for urban transport that incorporates most of the commonly seen strategies at a network level, one that considers partially observable user schedule disruptions. In this study, the strategy selection portion of transit incident management shown in Figure 1 is addressed with new models for more comprehensive strategy selection at a network level that accounts for duration uncertainty. Two models are tested over 20 different combinations of demand and disruption duration patterns. Comparisons with two other benchmark strategies are made.

The paper is organized as follows: Section 2 presents the hierarchical framework and two strategy selection models. Section 3 discusses the numerical tests on a toy example; two models are compared with two benchmark models. Section 4 concludes.

## 2 Methodology

*2.1 Framework*

In this study, we focus on metro systems. In the case of disruptions, efficient use of available resources is desired. Previous studies formulate this strategy selection problem for urban public transport the same way we handle intercity trains or airline systems: a service run is a basic unit of task. Train trajectories on the network are sought to cover these tasks. However, urban public transport system users typically do *not* buy tickets for a specific run or even a specific line. Instead, users pay for recurrent network services. The disruption mitigation is naturally separated into three phases: (i) Network level resource adjustment (i.e. the strategy selection); (ii) resource routing on the network; and (iii) local line adjustment. See Figure 1 for the activity diagram of hierarchical disruption mitigation. Modules corresponding to three phases run sequentially. The proposed models are implemented in Phase (i), where the strategy selection problem becomes a network resource allocation problem. Phases (i), (ii) and (iii) are run at network-level, regional-level, and line-level, respectively. That means multiple regions (or multiple lines) will run phase (ii) (or phase (iii)) concurrently.

We argue that the basic task unit to be adjusted for urban public transport resource allocation is best at a *line service level* instead of a service run level. Operators respond to disruptions by changing line service levels through diverting vehicles and crews between lines, including some newly setup *emergency lines*. Vehicles and crews may come from a high-cost backup depot. The service line level approach allows us to evaluate network-wide resources in a tractable manner while still accounting for user delays over a finite time horizon.

Two models are proposed in increasing levels of complexity that both include the line service level basic structure for network resource allocation. In Section 2.2, a novel "basic model" (BM) is first proposed to study the simplest deterministic disruption duration case. In Section 2.3, the BM is extended to a random model called "Initiation Time Model" (ITM). The two models differ in the way they handle random disruption duration (see Figure 2 for an illustration of the differences). BM uses expected values and treats duration as deterministic. ITM delays substantial actions to obtain further information over a single time horizon.

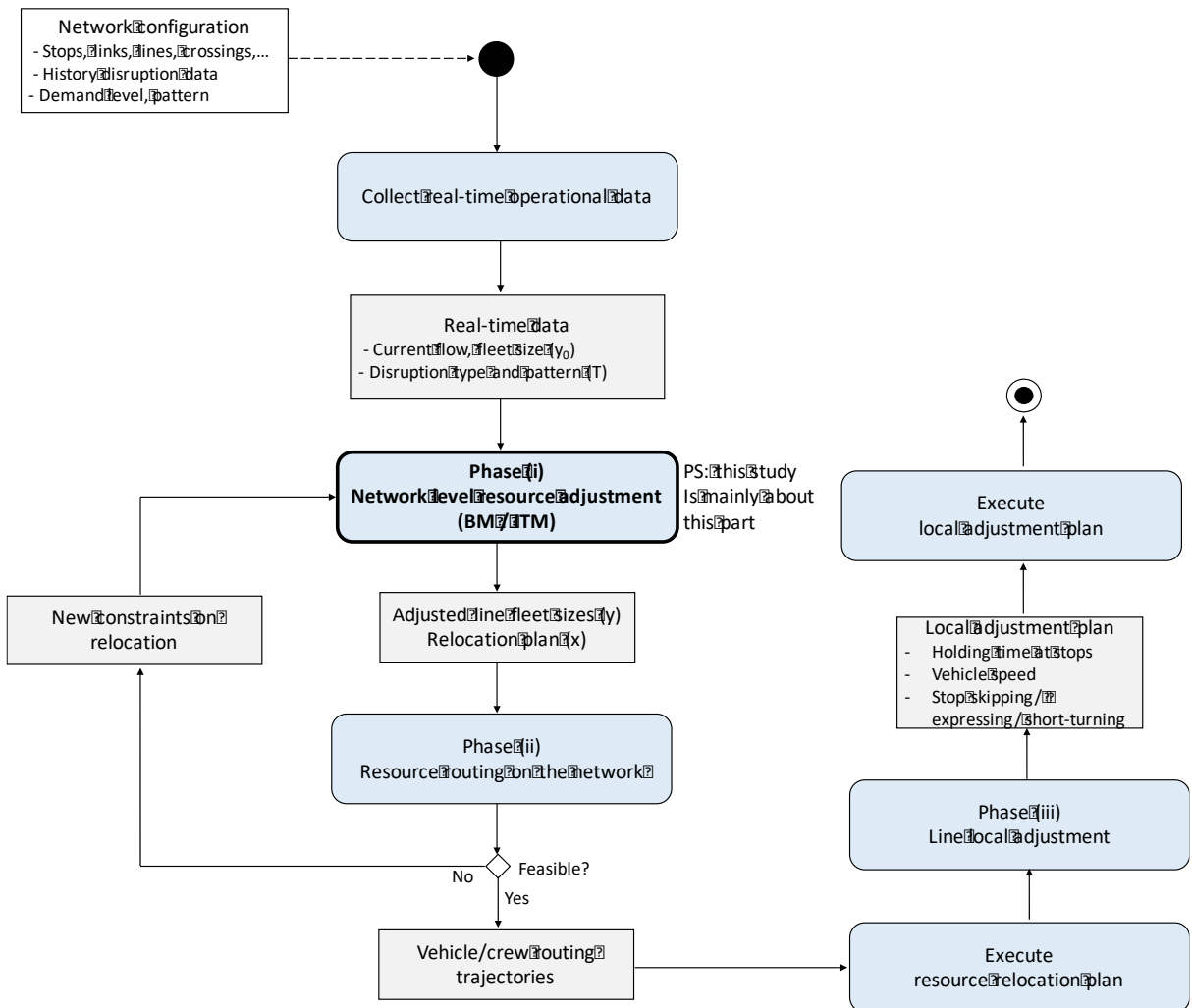

Figure 1. Activity diagram of hierarchical disruption mitigation.

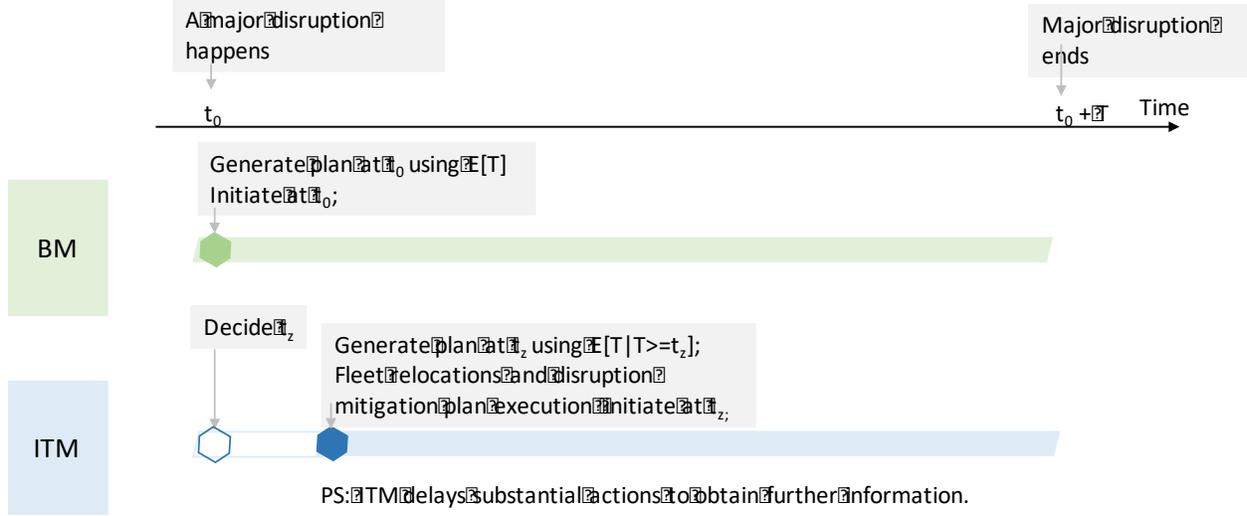

Figure 2. Comparison of BM and ITM.

## *2.2 Basic model (BM)*

*Notations*

$c_{ll'}$: average one-way cost of diverting vehicle from line $l$ to line $l'$ (constant);
$\mathbb{E}[T|T \geq z]$: expected duration conditioning on event $\{T \geq z\}$;
$f_l$: the frequency of line $l$;
$F^{BM}$: the objective of basic model (BM);
$F^{ITM}$: the objective of initiation time model (ITM);
$g(T)$: probability density function (pdf) of $T$;
$G$: transit network graph;
$H_w$: the set of paths between OD pair $w$;
$I^{ITM}$: ITM interval;
$L$: the set of transit lines;
$M$: set of transport modes;
$p_{w,h}$: the proportion of users of OD pair $w$ on path $h$ suggested by transit operator during disruption;
$p_{w,h}^N$: the path choices when system is undisrupted ("normal");
$p_{w,h}^D$: the path choices when system is disrupted and with no relocation ("disrupted");
$q_w(\tau)$: the user demand density for OD pair $w$ at time $\tau$;
$Q_w(t_1, t_2)$: the number of users belonging to OD pair $w$ during time interval $[t_1, t_2]$;
$Q_w$: the number of users belonging to OD pair $w$ during $[0, T]$; $Q_w \coloneqq Q_w(0, T)$;
$R_{l_s}$: round-trip time of line $l$ that is incident to segment $s$;
$S$: the set of transit line segments;
$S_h$: the set of segments on path $h$;
$S_h^B$: the set of boarding segments on path $h$;
$t_s^R$: running (traversing) time of transit segment $s$ (constant);
$t_h^P$: path $h$ cost during disruption after relocation finished;
$t_h^{P,N}$: path cost when system is undisrupted ("normal"), a constant;

$t_h^{P,D}$: path cost when system is disrupted and with no relocation ("disrupted"), a constant;
$t_h^P$: path cost when system is disrupted and with relocation;
$T$: disruption duration (a fix number) used in BM;
$\boldsymbol{T}$: disruption duration (a random variable) used in ITM;
$\bar{T}$: the upper bound of $\boldsymbol{T}$;
$V$: the set of transit stops;
$W$: the set of all OD pairs;
$x_{ll'}$: the number of vehicles relocated from line $l$ to $l'$ where $l$ and $l'$ are lines of the same mode $m \in M$;
$y_l$: adjusted fleet size for transit line $l$;
$K$: the capacity of the vehicle;
$y_l^0$: original line fleet size for line $l$;
$Y_l$: the upper bound of fleet size for line $l$;
$\mathcal{G}_s$: left hand side of Eq. (2);
$\mathcal{H}_l$: left hand side of Eq. (3);
$I$: left hand side of Eq. (4);
$\mathcal{J}_w$: left hand side of Eq. (5);
$\mathcal{K}_l$: left hand side of Eq. (6);
$\alpha$: weighing coefficient for operator cost;
$\beta$: user value of time (VOT) per minute;
$\gamma$: wait time penalty coefficient;
$\mu_s$: Lagrange multiplier for Eq. (2);
$\vartheta_l$: Lagrange multiplier for Eq. (3);
$\eta$: Lagrange multiplier for Eq. (4);
$\pi_w$: Lagrange multiplier for Eq. (5);
$\theta_l$: Lagrange multiplier for Eq. (6);
$\delta_{h,s}$: path $h$ and segment $s$ incidence;
$\delta_{h,l}$: path-line incidence;

Remarks:
1) Notation that appear only once are explained in text in place and not listed here;
2) A subscript is used for indexing, like '$w$' for OD pair, '$h$' for path, '$s$' for segment, '$l$' for line; superscript is used for differentiating, like '$B$' for 'boarding', '$R$' for running (traversing), '$D$' for diverting, '$P$' for path, '$0$' for naught.

The transit network is represented by a graph $G = (V, S)$ where $V$ is the set of transit stops and $S$ is the set of transit line segments. Initially, let disruption duration $T$ be a fixed real number. We assume that users follow paths. A *path* is composed of a sequence of transit line segments. This is a simplified version of user assignment under a disruption setting. Under steady state with limited real time transit information, users are assumed to be assigned to hyperpaths (Chriqui and Robillard, 1975, Spiess and Florian, 1989, De Cea and Fernández, 1993). In a disruption, however, we assume that the system can convey travel information to passengers and direct them to paths, eliminating the need for hyperpaths (a similar assumption is made by Mo et al., 2023). Let $W$ be the set of all OD pairs, indexed by $w$. Let $H_w$ be set of paths between OD pair $w$, indexed by $h$. Let $S_h$ denote the set of segments on path $h$. $S_h^B$ means the set of boarding segments on path $h$. A path may contain multiple boarding segments if a transfer exists. So, we have $S_h^B \subset S_h \subset S$. $L$ denotes the set of transit lines, indexed by $l$. The frequency of line $l$ is $f_l$. Figure 3 illustrates a transit network with two lines and one user traveling from stop A to E.

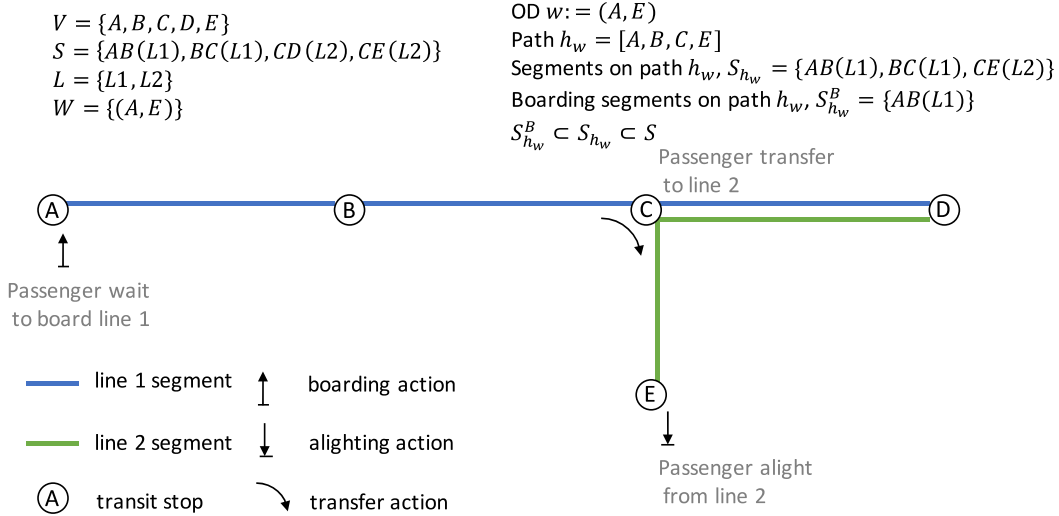

Figure 3. Transit network illustration (with the path of a transit passenger shown on the network)

There is no consensus about how users react to disruptions. In some studies, it is assumed that users are rational, self-interested, and always choose the shortest path (Cadarso et al., 2013). However, as argued by Xu and Ng (2020), under unforeseen disruptions, commuters may need to react with limited information. Instead, users can be guided toward alternative contingency routes by operators. As such, we let the user path choices be decision variables of the model (i.e. decisions of the operator). It is possible to assume that a proportion of the users comply with operator orders and the rest of them act on their own with full information. We leave this option for future work.

The novel line service level network resource allocation model is shown in Eq. (1) – (9). The decision variables are the fleet size for each line ($y$), assignment of users to paths ($p$), and the fleet relocation decisions among lines ($x$) needed to achieve $y$. Networks of different modes, like metro, bus, are jointly considered, where lines operate vehicles that belong to different non-interchangeable classes. In other words, $L = \bigcup_{m \in M} L_m$, where vehicles belonging to lines in class $L_m$ cannot be exchanged with vehicles in a line belonging to a difference class $L_{m'}$, $m \neq m'$. The variable $x_{ll'}$ is applied only to $l, l' \in L_m$.

*Formulation*

$$\min_{p,y,x} F^{BM}(p, y, x)$$

$$= \sum_{w \in W, h \in H_w} Q_w p_{w,h} \left( \sum_{s \in S_h^B} \frac{\gamma R_{l_s}}{2 y_{l_s}} + \sum_{s \in S_h} t_s^R \right) \quad (1)$$

$$+ 2\alpha \sum_{m \in M} \sum_{l,l' \in L_m} c_{ll'} x_{ll'}$$

Subject to:

(Segment capacity constraint)
$$\mathcal{G}_s := \sum_{w \in W, h \in H_w} Q_w p_{w,h} \delta_{h,s} - \frac{KT}{R_{l_s}} y_{l_s} \leq 0, \quad (\mu_s), \quad \forall s \in S \quad (2)$$

(Fleet size adjustments)
$$\mathcal{H}_l := \sum_{l'} x_{ll'} - \sum_{l'} x_{l'l} + y_l = y_l^0, \quad (\vartheta_l), \quad \forall l \in L \quad (3)$$

$$I := \sum_l y_l = \sum_l y_l^0, \quad (\eta) \quad (4)$$

(User path choices)
$$\mathcal{J}_w := \sum_{h \in H_w} p_{w,h} = 1, \quad (\pi_w), \quad \forall w \in W \quad (5)$$

(Fleet size bounds)
$$\mathcal{K}_l := y_l - Y_l \leq 0, \quad (\theta_l), \quad \forall l \in L \quad (6)$$

(Non-negativity)
$$p_{w,h} \geq 0, \quad \forall w \in W, h \in H_w \quad (7)$$
$$y_l \geq 0, \quad \forall l \in L \quad (8)$$
$$x_{ll'} \geq 0, \quad \forall l, l' \in L_m, m \in M \quad (9)$$

The objective is to minimize the weighted sum of costs to transit users and the operator (Eq. (1)) (see Zhang and Lo, 2018; Guo et al., 2019; Claessens et al., 1998; Cadarso et al., 2015). User cost is the trip time multiplied by value of time (VOT) $\beta$ as shown by the first term. Let $Q_w(t_1, t_2)$ be the number of users belonging to OD pair $w$ during time interval $[t_1, t_2]$. $Q_w(t_1, t_2) = \int_{t_1}^{t_2} q_w(\tau) d_\tau$, where $q_w(\tau)$ is the user demand density for OD pair $w$ at time $\tau$. Let $Q_w := Q_w(0, T)$. For those passengers that enter the system before the horizon begins, their location in the system at time 0 is regarded as their origins. We add $Q_w^0 \delta_0(t)$ to the density $q_w(\tau)$ to take account of these demands, where $Q_w^0$ with $w = (O, D)$ means the number of users queueing at $O$ heading to $D$ at time 0 and $\delta_0(t)$ is the Dirac delta function with a peak at $t = 0$.

We consider the user cost under stable flow condition. The complex process of transit system state transition during resource relocation is not modeled in this phase to avoid dynamic transit assignment modeling. The problem in reality is much more complex. As discussed in the literature on dynamic transit assignment (e.g. see Hamdouch and Lawphonpanich, 2008; Jin et al., 2016), time-varying travel times and flows mean that paths may not be easily categorized into pre-initiation/ recovery/ recovered stages. There could be passengers crossing the boundaries. Keeping track of these passengers will require the use of dynamic transit assignment with time-expanded networks (TE-network). The problem with adopting such frameworks is that they are not very scalable, which prevents the use of a strategy selection model at a network level. Instead, we try to keep things simple by assuming that the time horizon of the incident is small enough between those three stages that paths can be pre-identified for OD pairs. After selecting strategies, more detailed dynamic models may be deployed to aid implementation of the strategies in Phases ii and iii as shown in Figure 1.

The average user waiting cost of a boarding segment is computed by $\frac{R_{l_s}}{2y_{l_s}}$ where $l_s$ refers to the line of segment $s$ and $R_{l_s}$ is the roundtrip time of this line. The path $h$ has average cost $\sum_{s \in S_h^B} \frac{\gamma R_{l_s}}{2y_{l_s}} + \sum_{s \in S_h} t_s^R$ where $\gamma$ is the wait penalty coefficient and $t_s^R$ is the segment travel cost. The second term, operator cost, is the spending on resource relocation. Operator cost is weighted by a parameter $\alpha$. We assume that all fleet sizes restore to normality after disruption. $c_{ll'}$ is a unit one-way relocation cost. We do not restrict fleet size change variables $x$ and $y$ to be integral. The rounded values are typically good enough at a strategy selection level in phase i and can provide informative results for deploying strategies in phases ii and iii. Fractional results are not assumed to be in time, i.e. the duration would be for the full time horizon. For example, if a line has small fleet size, say $y = 0.3$, this line is typically an emergency line. In practice, an operator may round the value when implementing in phase iii. After all fractions are rounded, any violations to the feasibility constraints could be adjusted by judgment. We also allow $p$ to be fractional, which means the operator can control the exact proportion of users on a path. There are more sophisticated ways to estimate the passenger delay, like Sun et al. (2016). The use of fractions for passenger paths is even less of an issue than for frequencies, as passenger volumes tend to be high enough (e.g. rounding 287.8 to 288), just as all transit assignment models in the literature do not assume integer values.

Eq. (2) requires that the total demand to cross a *line segment* during $T$ be no larger than the expected capacity provided during $T$, which again depends on the average headway. Eq. (3) and Eq. (4) are about the fleet conservation constraints. Eq. (5) are the path flow conservation constraints and Eq. (6) are the fleet size bounding constraints. Eq. (7) are the non-negativity constraints. $\mathcal{G}_l, \mathcal{H}_l, I, \mathcal{J}_w, \mathcal{K}_l$ are functions representing left-hand side (LHS) of constraints; $\mu_l, \vartheta_l, \eta_l, \pi_l, \theta_l$ are the corresponding Lagrange multipliers of the constraints.

Paths are enumerated under this formulation. For convenience, k-shortest paths are used to approximate the true set of paths (see Bekhor et al. (2006)). While the appropriate number of $k$ paths to be chosen can vary with the size of the network (Bekhor et al., 2008), for simple networks Cascetta et al. (1997) showed that 4-7 paths may suffice.

Parameter $Y_l$ is the maximum fleet size of line $l$. $Y_l$ is determined by the throughput capability of line $l$. If multiple lines share some segments, the maximum fleet sizes of these lines are related; constraints like $\sum_{l_i=l_1}^{l_n} Y_{l_i} \leq c_s$ should be imposed. We leave this out in the formulation for simplicity. The relocation cost is defined in Eq. (10). The diverted fleets cannot provide regularly scheduled service during the diversion. This cost is captured by term $\gamma_{ll'}^D t_{ij}^D$. The costs of using backup vehicles and crews are represented by the term $\bar{c}_{ll'}$ when $l$ is a backup depot. $c^0$ represents the minimum costs associated with making diversions.

$$c_{ll'} = c^0 + \bar{c}_{ll'} + \gamma_{ll'}^D t_{ij}^D \tag{10}$$

where

$\gamma_{ll'}^D$: user cost for unit time spent on diverting unit vehicle from $l$ to $l'$ (which includes lost service on $l$ and unavailability on line $l'$ until its arrival);

$t_{ij}^D$: time that it takes to divert unit vehicle from line $l$ to line $l'$;

$\bar{c}_{ll'}$: vehicle and crew cost for diverting unit vehicle from $l$ to $l'$;

$c^0$: penalty for making changes.

BM is *generalized* in the sense that several other commonly seen models can be regarded as special cases. We use "fixing a network" to refer to the network topology being fixed but service level being subject to change; and use "fixing a service" to refer to both the network topology and the line service levels being fixed.
- Special case 1): If we fix the bus service, and allow metro network redesign as well as metro resource relocation, this is the *service line redesign problem*;
- Special case 2): If we fix bus service and fix the metro network, but can relocate the resources possibly across metro lines, this is the *service run adjustment problem*;
- Special case 3): If we fix the metro service but can adjust the bus services by adding bus-bridging lines and adjust bus line frequencies, this is the *substitution service design problem (bus-bridging problem)*;
- Special case 4): If we fix metro and bus network but can adjust the service levels of both original metro and original bus lines; this is the *multi-modal joint optimization problem*.

Figure 4 gives an illustration of the strategies considered in this study. When disruption happens, we can adjust current metro and bus line services, as well as set up new emergency metro and bus lines. The needed fleet could come from a backup depot or from existing lines. Not all strategies are needed at the same time. The best combinations are sought.

Note that a fraction of a fleet size (e.g. 3.76 vehicles) does not refer to operating only for a fraction of the whole disruption horizon. Rather, it refers to allocation of a portion of the fleet. In practice, this would be translated to rounded nearest integers with some local adjustments for edge cases.

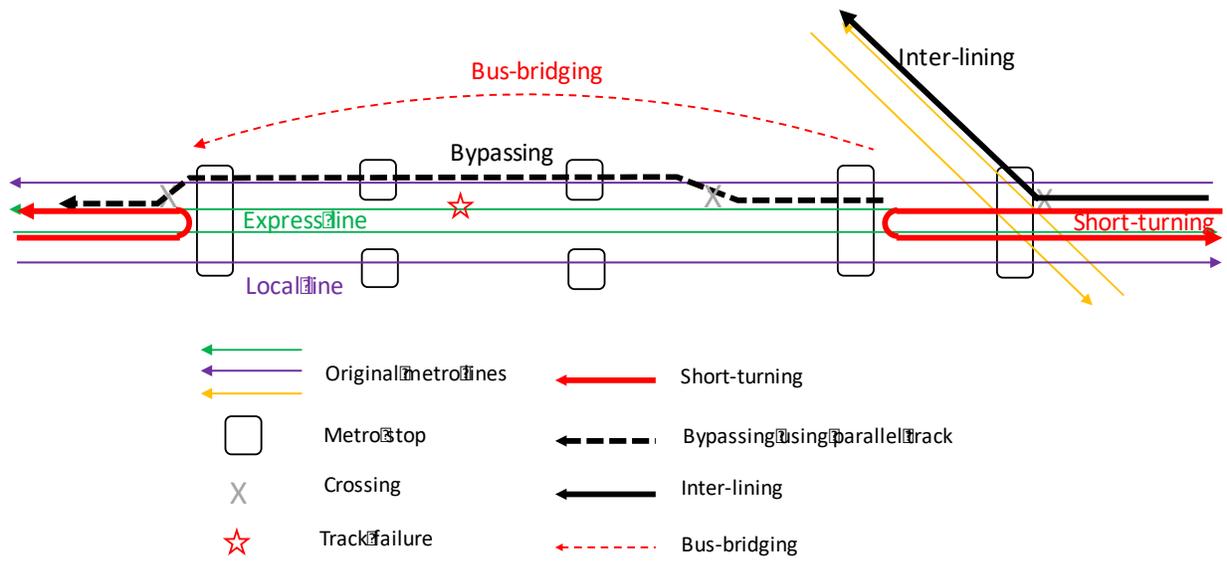

Figure 4. Strategies considered in this study

*Parameter estimation*
BM is parametrized by OD demands, cost coefficients, value of time, and the expected disruption duration. The strategy selection model in phase i assumes that the transit system knows at the start of the disruption which service lines are available, which are impacted, such that immediately available emergency lines can be determined instantaneously. Similarly, OD demand is assumed to be known. These assumptions are similar to the state of the art, as summarized in the literature review (see Section 1.4). For example, most modern transit systems have Advanced Vehicle Location (AVL) systems to keep track of their vehicle fleets at any time and can pinpoint the exact line segment or track section that is disrupted. Similarly, transit systems have historical data and Automated Passenger Counters (APC). Combined with state-of-the-art origin-destination inference methods (see Liu and Chow, 2023), transit systems can infer expected passenger OD flows over a time horizon. For example, NYC Transit keeps track of passenger arrivals through turnstile data and wifi detection using the TransitWireless system. They also have a transit control center that keeps track of the status of all rail segments in the subway system. These systems help provide a picture of passenger ODs and paths. Readers are recommended to follow studies on OD flow estimation (Castillo et al., 2015), network design problem for building set of emergency lines (Jin et al., 2016), and survival analysis for disruption duration (Tinguely et al., 2019) among others.

*Optimality conditions*
Eqs. (1) – (9) have a nonlinear objective with linear constraints. There are two weight coefficients $\alpha$ and $\beta$ in the objective. Without loss of generality, we may assume $\beta = 1$ after the transformation $\alpha = \alpha/\beta$. The KKT conditions are shown in Eq. (11).

$$\nabla \mathcal{L} := \nabla F^{BM} + \sum_s \mu_s \nabla \mathcal{G}_s + \sum_l \vartheta_l \nabla \mathcal{H}_l + \eta \nabla I + \sum_w \pi_w \nabla \mathcal{J}_w + \sum_l \theta_l \nabla \mathcal{K}_l \geq 0$$

$$\begin{aligned}
\nabla_{p_{w,h}} \mathcal{L} \cdot p_{w,h} &= 0, & \forall w \in W, h \in H_w \\
\nabla_{y_l} \mathcal{L} \cdot y_l &= 0, & \forall l \in L \\
\nabla_{x_{ll'}} \mathcal{L} \cdot x_{ll'} &= 0, & \forall l, l' \in L \\
\mathcal{G}_s \cdot \mu_s &= 0, & \forall s \in S \\
\mathcal{K}_l \cdot \theta_l &= 0, & \forall l \in L \\
\mu_s &\geq 0, & \forall s \in S \\
\theta_l &\geq 0, & \forall l \in L
\end{aligned} \quad (11)$$

along with primal constraints. From the KKT conditions, we have the following observations.
**Observation 1**. Condition for path $h$ belonging to OD $w$ to be in use is
$$p_{w,h} > 0 \Rightarrow \nabla_{p_{w,h}} \mathcal{L} = 0$$
$$\nabla_{p_{w,h}} \mathcal{L} = Q_w t_h^P(y) + \sum_s \mu_s Q_w \delta_{h,s} + \pi_w = 0$$
Where path cost $t_h^P(y) = \sum_{s \in S_h^B} \frac{\gamma R_{l_s}}{2 y_{l_s}} + \sum_{s \in S_h} t_s^R$;
or, equivalently, in Eq. (12).
$$t_h^P(y) + \sum_s \mu_s \delta_{h,s} = -\frac{1}{Q_w} \pi_w \quad (12)$$

The first term on the LHS is the user cost of path $h$; the second term on the LHS are segment-capacity shadow prices. The RHS can be interpreted as the cost of the marginal shortest path for OD $w$: the cost of sending marginal flow along the shortest path under the optimally loaded flow. Note $\pi_w$ is unrestricted. This condition says that, if a path $h$ for OD $w$ is used, then its cost plus the segment-capacity shadow prices equals the marginal shortest path length. This type of condition is common for a multicommodity flow problem.

**Observation 2**. Condition for emergency line $l$ to be in use is
$$y_l > 0 \Rightarrow \nabla_{y_l} \mathcal{L} = 0$$
$$\nabla_{y_l} \mathcal{L} = \sum_{w \in W, h \in H_w} \sum_{s \in S_h^B, l_s = l} \left( -\frac{\gamma R_l}{2 y_l^2} \right) - \mu_l \frac{KT}{R_l} + \vartheta_l + \eta + \theta_l = 0$$

Moving some negative terms to the RHS, we get:
$$\vartheta_l + \eta + \theta_l = \left( \sum_{w \in W, h \in H_w} Q_w p_{w,h} \sum_{s \in S_h^B, l_s = l} \frac{\gamma R_l}{2 y_l^2} \right) + \sum_{s \in S_h^B, l_s = l} \mu_s \frac{KT}{R_{l_s}} \tag{13}$$

$\vartheta_l$ is the multiplier associated with relocation flow $x$ conservation; it is the node potential in the transportation problem. It could be interpreted as the marginal cost of diverting vehicles to line $l$. $\eta$ is the shadow price of fleet resource. $\theta_l$ is the price associated with upper bound $Y_l$ which could be positive if line $l$ is operated at capacity. The first term on the RHS is the (positive) waiting time savings of users with respect to unit $y_l$ increase. The second term on the RHS is the marginal benefit of improving line $l$ capacity which could be nonzero if some segment belonging to $l$ operates at capacity. Hence the equation means marginal cost of diverting plus fleet shadow price and fleet upper bound shadow price are equal to the marginal savings of user wait time plus marginal benefits of expanding capacity. Conversely, if the following condition is satisfied, then we must have $y_l = 0$; namely, this emergency line is not in use.

$$\vartheta_l + \eta > \left( \sum_{w \in W, h \in H_w} Q_w p_{w,h} \sum_{s \in S_h^B, l_s = l} \frac{\gamma R_l}{2 y_l^2} \right) + \sum_{s \in S_h^B, l_s = l} \mu_s \frac{KT}{R_l} \tag{14}$$

**Observation 3**. Condition for fleets being diverted from line $l$ to line $l'$ is
$$x_{ll'} > 0 \Rightarrow \nabla_{x_{ll'}} \mathcal{L} = 0$$

$$\nabla_{x_{ll'}} \mathcal{L} = 2 \alpha c_{ll'} + \vartheta_l - \vartheta_{l'} = 0 \tag{15}$$

where $2 \alpha c_{ll'}$ is the cost of $x_{ll'}$; $\vartheta_l$ is the node potential as we mentioned. This is exactly the optimality condition for a transportation problem.

*Solution method*
The BM formulation can be generalized to (P0), which has potential for broader applications. Variable $p$ is the user demand assignment decision; $y$ is the service level decision for lines (or any other type of service entities); $x$ is the resource diversion decision among lines. The objective is composed of the diversion cost and user cost. It has nonlinear terms like $\frac{p_{w,h}}{y_l}$ in the objective which represent delay from a deterministic queueing system. This objective is not convex. For other types of queueing systems, the exact formulation may be different, but what is in common is nonconvexity. Take M/M/1 for example; average delay has the form of $\frac{v}{c-v}$ where $v$ is link flow and $c$ is capacity; when $v$ and resource $c$ are decision variables, this delay function is also not convex. Here the constraints are Eq. (2) to (9) as before, although other types of systems may call for changes. We call (P0) the nonconvex Joint Routing and Resource Allocation (nJRRA) problem. This problem shares similar properties with the multicommodity capacitated network design problem, which differs in the use of binary variables to allocate link investment resources while subject to optimal passenger flows (see Gendron et al., 1999).

(P0)

$$\min F(p, y, x) = c'x + c''p + \sum_{w,h,l} c'''_{w,h,l} \frac{p_{w,h}}{y_l}$$

s.t. Linear constraints (2) – (9)

Convex or nonconvex JRRA problems arise when studying many different types of networks, like transit networks, computer networks, or power grids. Operators (or ISPs for internet, utility companies for power grids) plan the resource relocation and can control how flows are distributed on the network at the same time. Xiao et al. (2004) studied the JRRA problem (called "simultaneous routing and resource allocation (SRRA)" there) for a wireless network. They assumed the objective to be convex for minimization and concave for maximization, like the utility function. The problem is solved through Lagrange duality. Capacity multipliers are introduced, then the resulting Lagrange dual problem can be decomposed. A subgradient method is used to update capacity multipliers. El-Sherif and Mohamed (2013) studied JRRA minimizing delay for cognitive radio based wireless mesh networks. The objective is similar to (P0). Their model is formulated as mixed integer programming. Similar studies include Rasekh et al. (2019). In this section, we discuss global solution algorithms for nJRRA.

The domain is compact. Note $y_l = 0$ is within the domain. We define $\frac{p_{w,h}}{y_l}$ to be 0 if $p_{wl}$ and $y_l$ are both zero. Namely, if a line has no vehicle, and if no user is diverted to this line, then the user cost accumulated on this line is zero. If $y_l = 0$ for $l$ and $p_{w,h} > 0$ for some $w$ and $h$ and path $h$ uses this line $l$, then objective $F$ of (P0) becomes infinite. So, the dependence of $F$ on $p$ and $y$ is discontinuous at $y_l = 0$. The logical relation in Eq. (16) holds at optimality but the reverse is wrong in general.

$$\left(y_l^* = 0 \land \delta_{h,l} = 1\right) \Rightarrow p_{w,h}^* = 0, \forall w \tag{16}$$

The essential singularity point at the boundary caused by $p/y$ terms may bring trouble to the convergence of iterative algorithms. Hence, we define a more constrained version of P0 that additionally requires $y_l$ to be no less than a small positive number $\epsilon$, say 0.01. If we find that the algorithm outputs $y_l^* = \epsilon$, then we can safely regard $y_l^*$ as zero for practical purpose. Let $u_l \coloneqq \frac{1}{y_l}$, then $u_l$ is bounded above by $1/\epsilon$. In this way, our problem has a compact domain and the objective is smooth on this domain. With new variable and new constraints added, we have problem (P1) reflecting an "$\epsilon$-constrained nJRRA". As $\epsilon \to 0$, P1 approaches P0.

(P1)
$$\min cx + c'p + \sum_{w,h,l} c'''_{w,h,l} p_{w,h} u_l$$
s.t. Linear constraints in (P0)
$$u_l y_l = 1, \quad \forall l$$
$$\epsilon \leq y_l \leq \bar{y}_l, \quad \forall l$$
$$\frac{1}{\bar{y}_l} \leq u_l \leq \frac{1}{\epsilon}, \quad \forall l$$

The $\epsilon$-constrained JRRA problem is a special case of a Quadratically Constrained Quadratic Program (QCQP), although not every QCQP is of type (P1) and there may exist more efficient algorithms dedicated to the nJRRA problems. This is reserved for future research.

QCQP with a nonconvex objective is generally NP-hard (Pardalos and Vavasis, 1991). QCQP is a fundamental problem that has been studied extensively in the global optimization literature; a partial list of recent studies includes Al-Khayyal et al. (1995), Audet et al. (2000), Linderoth (2005), Qu et al. (2008), Zheng et al. (2011), Misener and Floudas (2012), Anstreicher (2012), Mitchell et al. (2014), Zhao and Liu (2017), Elloumi and Lambert (2019), Alkhalifa and Mittelmann (2022).

The number of OD pairs and paths to be considered by the model can be huge, thus making the model difficult to solve. For practical purpose, operators can restrict themselves to consider only:
- representative OD pairs whose demands are significant;
- OD pairs with users likely to be impacted by the disruption;
- representative paths.

A program should be designed to impose these restrictions automatically.

Two limit cases are discussed below to draw insights on the BM strategy.

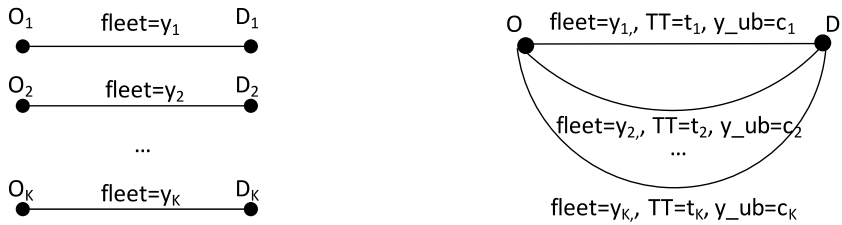

(a) Resource relocation between independent lines; The optimal solution is *square-root rule*:
$$y_1 : y_2 : \ldots : y_K = \sqrt{q_1} : \sqrt{q_2} : \ldots : \sqrt{q_K}$$

(b) Resource relocation between parallel lines; The optimal solution is *shortest path first rule.*

Figure 5. Two special cases.

For the first case (Figure 5 (a)), assume that there are $K$ lines connecting $K$ different OD pairs and there is no user interaction of any kind. Also, assume that we can ignore the relocation cost, namely $x$'s have coefficient zero. This simplified version of the problem can be written as:
$$\min_y \sum_k \frac{q_k}{y_k}$$

$$\sum_k y_k = c$$
$$y_k \geq 0$$

where
$q_k$: demand of line $k$;
$y_k$: feet size of line $k$;
$c$: total number of vehicles.

We can easily solve by using the first order conditions to find that at optimality, Eq. (17) holds:

$$y_1 : y_2 : \ldots : y_K = \sqrt{q_1} : \sqrt{q_2} : \ldots : \sqrt{q_K} \tag{17}$$

This corresponds to the *square root rule* (Furth and Wilson, 1981) – the fleet sizes should be proportional to the square roots of number of passengers. As for the second case (Figure 5(b)), suppose there are $K$ lines connecting one single OD pair. Each line $k$ has travel time $t_k$ and fleet size upper bound $c_k$. Also, suppose $t_1 < t_2 < \cdots < t_K$. The optimal solution is obvious: first assign fleets of size $y_1 = \min\{c_1, c\}$ to line 1; If there are vehicles left, then assign $\min\{c_2, c - y_1\}$ to line 2; continue until we run out of vehicles. We also give a name to this simple strategy – *shortest path first rule*.

## 2.3 Initiation time model (ITM)

When a disruption happens, it may be advantageous for an operator to wait for some time before taking any costly actions like bus bridging or inter-line vehicle diversion. Ideally, disruption mitigation should be modeled as a continuous decision-making process. For simplicity, the ITM model assumes that vehicle relocation initiates only once in the horizon. The exact time to start such a relocation is up to the operator. If the disruption recovers while waiting, then there is no need to make any relocations. Delaying actions reflects the principle that there is a tradeoff between the user cost and operator cost. This idea can be found in Zhang and Lo (2018), although they only focus on a single disrupted metro line and a single strategy: bus bridging.

Let $T$ now be the random disruption duration. Suppose $T$ is bounded above by $\bar{T}$ and suppose that it is continuously distributed with probability density function (pdf) $g$. We add a new variable $z$ - the relocation initiation time. The other variables are the same as before and the problem is labeled (P2). The objective (Eq. (18)) has three terms. The first term corresponds to the user cost when $T < z$, the case that the relocation has never been initiated. The second term corresponds to the user cost when $T \geq z$. The second term can again be decomposed into three sub-terms, corresponding to pre-initiation, recovery, and recovered periods. The expected operator cost is captured by the third term. Eq. (19) means that the capacity in the interval $[z, \mathbb{E}[T|T \geq z]]$ can satisfy the demand in that period where $\mathbb{E}[T|T \geq z]$ means $T$ is conditioned on the event that the disruption has not ended at time $z$. Eq. (20) is the upper bound on $z$ and $\rho$ is its corresponding Lagrange multiplier.

*Formulation*

(P2)

$$\min_{z,p,y,x} F^{ITM}(z,p,y,x) =$$

$$\underbrace{\sum_{w \in W, h \in H_w} \int_0^z \left( Q_w(0,T) p_{w,h}^D t_h^{P,D} + Q_w(T,\bar{T}) p_{w,h}^N t_h^{P,N} \right) g(T) dT}_{\text{case } T<z: \text{ user cost in } [0,\bar{T}]}$$

$$+ \underbrace{\sum_{w \in W, h \in H_w} \int_z^{\bar{T}} \left( \underbrace{Q_w(0,z) p_{w,h}^D t_h^{P,D}}_{\text{user cost in } [0,z]} + \underbrace{Q_w(z,T) p_{w,h} t_h^P(y)}_{\text{user cost in}[z,T]} + \underbrace{Q_w(T,\bar{T}) p_{w,h}^N t_h^{P,N}}_{\text{user cost in}[T,\bar{T}]} \right) g(T) dT}_{\text{case } T \geq z: \text{ user cost in } [0,\bar{T}]} \quad (18)$$

$$+ \underbrace{2\alpha \sum_{l,l'} c_{ll'} x_{ll'} \int_z^{\bar{T}} g(T) dT}_{\mathbb{E}[fleet\ size\ adjust\ cost]}$$

(Segment capacity constraint)

$$\mathcal{G}_s := \sum_{w \in W, h \in H_w} Q_w(z, \mathbb{E}[T|T \geq z]) p_{w,h} \delta_{h,s} - \frac{K}{R_{l_s}} (\mathbb{E}[T|T \geq z] - z) y_{l_s} \leq 0, \quad (19)$$
$$(u_s) \; \forall s \in S$$

$$\mathcal{R} := z - \bar{T} \leq 0, \quad (\rho) \quad (20)$$

Subject to Eq. (3) to (9).

*Solution method*

The capacity constraints are nonlinear now. The rest of the constraints (Eqs. (3) - (9)) are linear as before. The objective is nonlinear and more complex than that of BM. We note that if we fix $z$, the solution algorithm previously discussed still applies with some minor changes. It is usually more convenient to discretize the time for application. Here we describe a practical way to speed up discrete ITM; we call it *early-break-ITM* (Algorithm 2). First, we discretize $\bar{T}$ and restrict the candidate initiation time to be a multiple of an ITM_interval, $I^{ITM}$. The idea is to start with $z = 0 \times I_{ITM}$ and increase $z$ until no successive improvement can be made; then break out of the iteration and return the last $z$. We call the sub-problem of ITM with fixed variable $z$ by the name (P3).

**Algorithm 2: Early-break-ITM**

Start with $z = 0$, $z^{opt} = \emptyset$, $F^{opt} = \infty$;
While $F^*(z) < F^{opt}$:
    $z^{opt} = z$;
    $F^{opt} = F^*(z)$;
    $z = z + I^{ITM}$;
    Solve (P3) to get $F^*(z)$;
Output $z^{opt}$ and $F^{opt}$.

Problem (P3) has the same complexity as BM. From our experiences, delay time $z$ is mostly within 30 minutes. In Algorithm 2, if $I_{ITM}$ is set to be 10 minutes, four iterations would suffice in most cases. Hence, the complexity of Algorithm 2 would typically be about four times that of the BM solution algorithm.

## 3 Numerical Tests

### *3.1 Evaluation program design*

Since the models need to be evaluated, a quasi-dynamic evaluation program is used (since the demand is deterministic, we do not call it a simulation). Note that the line capacity and transit flow remain static (i.e. there's no spillback, individual vehicle runs or vehicle capacities). The evaluation program has a main function and five modules: network generation, demand generation, disruption generation, mitigation plan generation, and evaluation module. Network generation is responsible for generating network components, producing lists of stops, transit lines, links, paths, as well as path-link incidence matrix. Paths are enumerated for the simple network to be introduced. The number of reasonable paths between an OD pair can be as much as six.

The program runs network, demand, and disruption generation functions in sequence. Then the disruption mitigation strategy is generated at the beginning of the horizon. The program loops over time and evaluates the expected user costs. Expected user costs are accumulated as shown in Eq. (21). Note that we assume the simpler case of demand being known and deterministic, hence the only stochastic factor is the disruption duration. Therefore, we can always generate the strategy for the whole horizon from the beginning. Under our assumptions, the returned strategy includes all the information that an operator needs to know to act when the uncertainty of disruption unfolds in the future. More complex models considering stochastic demand along with the corresponding simulation are possible and reserved for future work. The disruption ending time is simulated according to disruption generation function. The program is written in Python, with optimization via Gurobi 8.1.1 API. The program is run on a Dell personal computer with Intel Core i5 and 8GB DDR3 memory. The details of the evaluation program are shared at https://github.com/BUILTNYU/transit-disruption-mitigation/.

$$F^{SIM} = \sum_{T=1}^{\bar{T}} \left( g(T) \sum_{k=1}^{\bar{T}/I^{SIM}} \left( I^{SIM} \sum_{w \in W} N_{T,k,w} \right) \right) \qquad (21)$$

We test the proposed models on a toy transit network for reproducibility (Figure 6). The network has two metro lines and two bus lines (Figure 6 (a)). Each horizontal and vertical link is assumed to have travel time of 8 minutes for bus and 4 minutes for metro. Diagonal links cost 6 minutes for metro. The round-trip time is 36 minutes for metro line and 96 minutes for bus line. The total number of trains is 6; the total number of buses is 36 (32 in use and 4 for back-up). Each train can carry 800 passengers and each bus can carry 100; these values are used in combination with the frequencies to obtain line capacities.

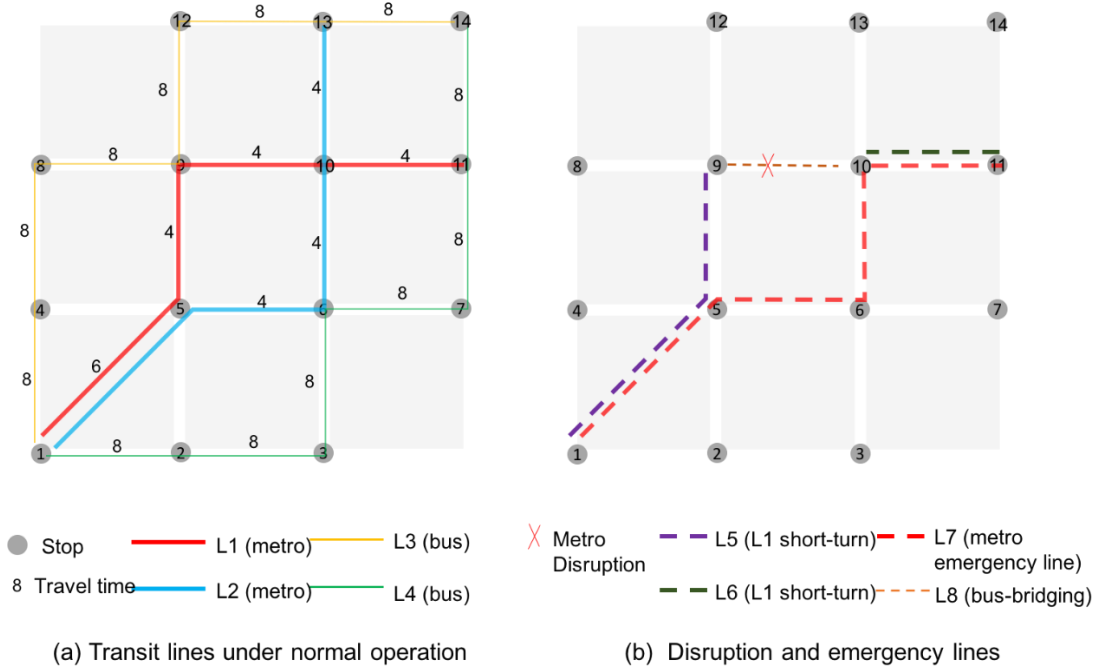

Figure 6. Example network and disruption

The main parameters of our models include the value of time (VOT), the operational cost coefficients, demand pattern coefficients and disruption shape coefficients. There are extensive studies on VOT estimation. Operational cost coefficients are likely to differ from one system to another and should be calibrated carefully by each practitioner. Demand pattern and disruption shape coefficients are easy to calibrate by using historical demand and disruption records.

### 3.2 Deterministic disruption duration case

Eight representative OD pairs are considered: 1-10, 5-14, 3-13, 8-11, 11-2, 13-4, 14-1, and 10-5. The time-dependent demands are assumed to be deterministic and concave, represented by parabolic functions. We will use two parameters $q_{min}$ and $q_{max}$ to specify the curve (Figure 7):

$$q(t) = -\frac{4}{T^2}(q_{max} - q_{min})t^2 + \frac{4}{T}(q_{max} - q_{min})t + q_{min}$$

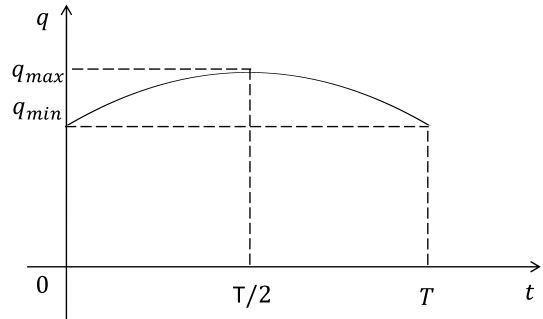

Figure 7. User demand pattern (parabolic)

The disruption to be considered is the failure of bi-directional link N9-N10 on line L1, say, due to tunnel power failure. The disruption lasts for $T = 60$ minutes for certainty. Four emergency lines—L5, L6, L7, and L8—are generated manually for this example as shown in Figure 6 (b). For real networks, the generation of these candidate lines should be automated and reserved for future research. L5 and L6 are short-turned metro lines for L1. L7 is a detour of the broken link in L1 using the L2 track. L8 is a bus-bridging line connecting metro stops 9 and 10.

**Parameter summary**
$T = 60$ minutes  # deterministic disruption duration
$q_{min} = 10$  # demand density parameter
$q_{max} = 12.5$  # demand density parameter
BLT = 100  # bus line transfer cost ($)
BBT = 300  # bus back-up transfer cost ($)
MLT = 200  # metro line transfer cost ($)
MST = 0  # metro short-turn cost ($)
num_metro = 6  # the total number of metro vehicles
num_bus = 36  # the total number of buses
capacity_metro = 800  # capacity of each train
capacity_bus = 100  # capacity of each bus
VOT = 0.1  # value of time ($) per minute
α = 1  # weighting coefficient for operator cost

For this deterministic case, three models are considered: line-level adjustment (LLA), bus-bridging (BB) and basic model (BM). LLA includes line level strategies, like short-turn and diverting users, but there is no inter-line fleet exchange. BB allows any strategies in LLA, and also allows the operator to allocate buses from a backup depot or existing lines to bridge the disputed links. BM allows any strategy in LLA and BB, and allows fleet exchange among different lines. We can see that BB is an extended model of LLA and BM is extended model of BB.

The convergence rate of BM is illustrated in Figure 8. The gap is defined as $(UB - LB)/UB$ where $LB$ and $UB$ are notations from Algorithm 1. Note that the "gap" measures how far the current best solution's objective is from the current relaxation, not the distance to the optimum. As such, even an optimal solution may have a nonzero gap. Within 2 minutes, the gap drops from an initial value of 20% to 10%. In 10 minutes, it drops to 8%. Afterwards, the trajectory becomes quite slow as there are many leaves on the branching tree. This convergence behavior is typical in such algorithms for QCQPs.

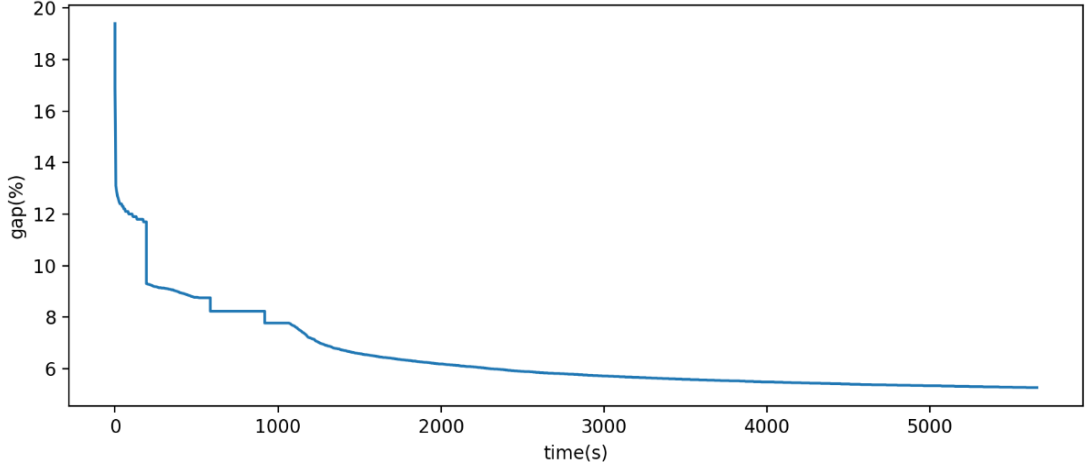
Figure 8. Gap changes over iterations (BM model solved by Gurobi).

The testing results are shown in Table 1. BM runs for 5 minutes, timing out before converging. LLA and BB run much faster (solved within 1 minute) and their results represent global optima (since each problem has a different objective, we do not expect the values to be equal). BM, even with a suboptimal solution, has almost the same level of service for users compared with an optimal BB, but at a much smaller operator cost. The improvement may not seem impressive at first sight, just about 4 percent compared to BB; but note that the "total costs" include the costs of all users on the whole network, disrupted or not. In summary:
- Introducing more strategies can significantly reduce operational cost without compromising user service levels.

This result supports the use of comprehensive strategy selection models instead of models that focus on single strategy optimization.

Table 1. Performance comparisons under deterministic disruption duration case

| Model | User cost ($) | Operator cost ($) | Total cost ($) |
|---|---|---|---|
| Line-level adjust (LLA) | 16607.5 (100%) | 0 | 16607.5 (**100%**) |
| Bus-bridging (BB) | 15237.5 (91.2%) | 1200 | 16437.5 (**98.9%**) |
| Basic model (BM)* | 15260.9 (91.9%) | 353 | 15614.2 (**94.0%**) |

* Algorithm timed out at 5 minutes without converging.

### 3.3 Stochastic disruption duration case

Next, we test BM and ITM with stochastic disruption duration $T$. The maximum duration of disruption, $\bar{T}$, is set to be 4 hours. We evaluate demand and disruption in a continuous way by specifying the demand density and disruption duration distribution pdf. However, this requires us to customize the models for each pattern where the integration of demand density and disruption pdf are involved. Some distributions may not have an explicit form of cumulative distribution function, i.e. the normal distribution. Hence, we represent demand and disruption temporal distributions with finite dimensional vectors. Assuming a time interval for demand and disruption to be 10 minutes, demand and disruption duration distribution are represented by vectors of dimension $\bar{T}/10$. Demand rates are assumed to be flat within an interval. Disruption duration is assumed to be in multiples of this interval. Note this interval does not have to be the same as the evaluation platform interval. To apply BM to a stochastic duration case, the expected value of disruption is used.

*Demand patterns*
Five different demand patterns over time are used: uniform, increasing, decreasing, concave and convex (Figure 9). They are represented by zero, first and second order polynomials. The concave and convex functions are similar to what we used in the deterministic case. All these patterns have two range parameters $q_{max}$ and $q_{min}$. When the demand pattern is uniform, the density is $1/2(q_{min} + q_{max})$.

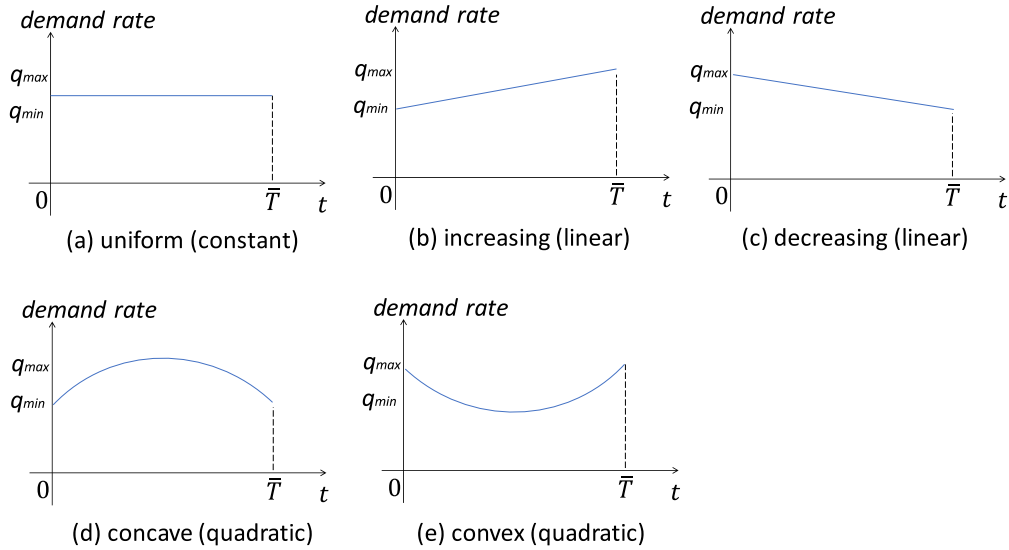

Figure 9. Five demand patterns

*Disruption duration distribution*
Probability mass functions (pmfs) are used to reflect the probability of the disruption duration $\boldsymbol{T}$. Six disruption duration distributions are illustrated in Figure 10: Dirac at time zero ($\delta_0$, "Dirac-0"), Dirac at $\bar{T}$ ($\delta_{\bar{T}}$, "Dirac-Tub"), weighted sum of Diracs at time 0 and $\bar{T}$ ($1/2(\delta_0 + \delta_{\bar{T}})$, "bi-Dirac"), uniform, "normal-like", and "exponential-like". Note that Dirac-0 means that the disruption lasts for one time interval (10 min). The first three distributions are simple and interesting for theoretical purposes. The first two are even deterministic. They can be treated as limiting cases of more complicated patterns to be considered. We list these three simple distributions since they can help us to understand the problem.

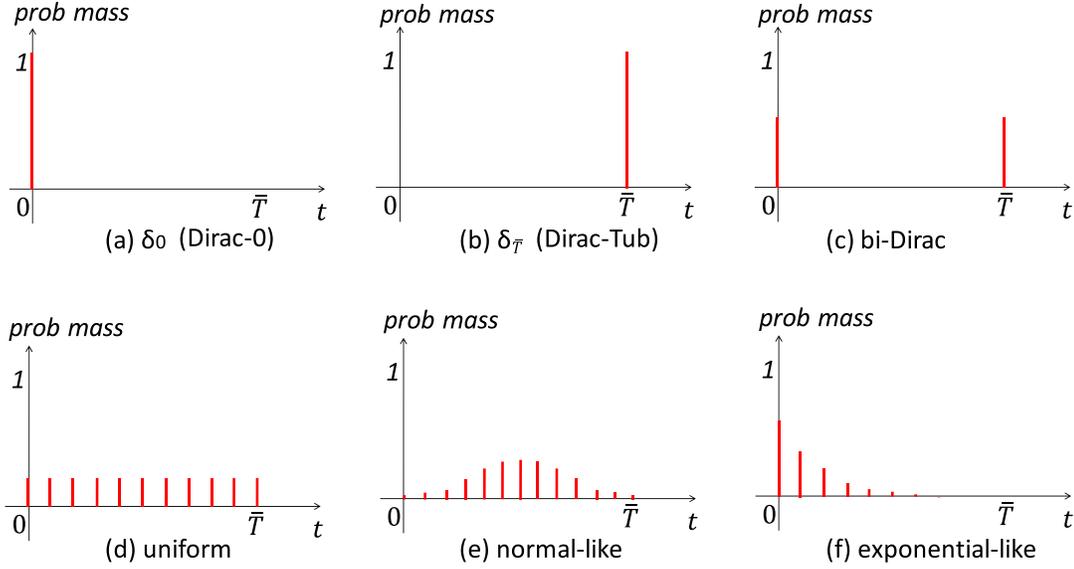

Figure 10. Six disruption distributions

When the duration distribution is $\delta_0$, if the time interval is small enough, we expect that all models will choose to make no fleet size change. Namely, the global optimums of these models are the same:
$$F^*_{LLA} = F^*_{BB} = F^*_{BM} = F^*_{ITM}$$

When the duration distribution is $\delta_{\bar{T}}$, we expect that all models will choose to do the best they can immediately. The following relationship holds:
$$F^*_{LLA} \geq F^*_{BB} \geq F^*_{BM} = F^*_{ITM}$$

When the duration distribution is a Dirac with a mass somewhere between 0 and $\bar{T}$, we expect to get results between two extreme cases above. This case itself could be considered the limiting case of a Normal-like distribution. We consider a bi-Dirac distribution for theoretical purposes, although it seems unlikely to occur in practice. A Bi-Dirac distribution means that the disruption will either stop very soon or will last as long as it could. It could be considered the limiting case of a bi-modal distribution. We expect ITM to postpone adjustments to time 2; and this is verified by the tests later.

The uniform distribution is another theoretically important distribution, though not quite likely to occur. For practical purposes, the two most important distributions are normal-like and exponential-like (truncated geometric), which correspond to Normal and exponential distributions in continuous modeling. For the latter three disruption distributions, it's not intuitively clear how they will perform in the tests.

*Test results*

We test five demand patterns and four disruption duration distributions (uniform, normal-like, exponential-like, bi-Dirac), 20 combinations in total. The detailed results are shown in Table A1 through Table A5 in the Appendix. "Operator cost" is part of the output of the mitigation plan generation model. "# BU bus" in the table means the number of back-up buses used; "z" means the initialization time found (min). Note the models are not solved with the same objective: LLA,

BB and BM uses Eq. (1); ITM uses Eq. (18). Each model runs for 5 minutes. The algorithms do not necessarily converge to the global optimum under the 5-minute running time constraint (runs that time out at 5 minutes are noted in appendix tables). However, they are all evaluated using the same metric, "Expected User cost" (Eq. (21)); both the objective values and the performance metrics are reported in the tables. The operator cost weighting factor alpha is set to be 2. ITM interval is set to be 10 minutes, hence there are 24 candidate initiation time points. Demand level is set at $q_{min} = 10$ and $q_{max} = 20$. All $\bar{T}/10$ possibilities of disruption duration are tested and the resulting expected costs are determined under uncertain disruption duration. Parameter changes are summarized below.

**Parameter summary**
$\bar{T} = 240$ minutes
$q_{min} = 10$
$q_{max} = 20$
time_interval = 10 min
ITM_interval = 10 min

Based on the test results, some observations are made below. We label cases by "demand pattern, duration distribution", like "uniform, normal-like".

**Remark 1**. *The overall performance of ITM is the best compared to LLA, BB, and BM among the instances tested.*

ITM outperforms BM significantly when the duration distribution is bi-Dirac; the overall performance of BM is otherwise very close to ITM. When initiation time $z$ is zero, the ITM model performance is slightly behind BM, like the case "uniform, uniform" in Table A1. This is because ITM is more difficult to solve, and its gap is larger for the 5-minute running time. For cases in which $z$ is positive, ITM is significantly better than BM, like the case "uniform, bi-Dirac" in Table A1.

**Remark 2**. *The overall performances of LLA and BB are significantly worse than that of BM and ITM.*

**Remark 3**. *When the demand pattern is concave or the duration distribution is bi-Dirac/exponential-like, it is advantageous to postpone the resource relocation decision.*

For example, for the case "increasing, exponential" in Table A2, ITM generates $z = 20$.

**Remark 4**. *When the demand pattern tends to be uniform or even decreasing, which means most of the users will arrive in the near future, or when disruption is likely to last for a long time, it makes less sense to delay actions.*

As we can see from Table A1 and Table A3, as long as the duration distribution is not bi-Dirac, ITM delay $z$'s are zero.

**Remark 5**. *When the disruption distribution is exponential-like, a backup bus is not used in BB.*

For cases with "exponential-like" disruption distribution in Table A1 to Table A5, the number of backup buses used by the BB model is 0. Backup buses are heavily used when the demand pattern is concave or decreasing. For cases where the demand pattern is "concave" and the disruption distribution is not "exponential-like", the number of backup buses used is 2.

**Remark 6**. *ITM will not delay for more than an upper limit due to the penalties of user delay.*

In the instances tested, ITM never delays more than 30 minutes.

*Sensitivity tests for alpha*

The effect of weighting parameter $\alpha$ is also investigated. The overall effect is that as $\alpha$ gets larger, the operating cost becomes more important, the number of relocations decreases, and the adjustments are initiated later. These effects match our expectations. This is illustrated by the case 1 to 3 in Table 2. We just listed two indices - the number of backup buses used by BB and initialization time of ITM; they are enough to provide insights on the effects of alpha. The detailed results are shared in Table 2. We also notice that there are many cases in which the relationship between the number of relocations and alpha is not so obvious. Cases 4 to 6 in Table 2 illustrate this situation. Take case 4 for example. When alpha increases from 10 to 20, ITM initialization time decreases from 30 to 10. This is counter-intuitive: why would operator initialize relocations earlier when they care more about their own costs? Looking closer at the results we find that the number of relocations is less - the number of backup buses used in BB is less than before.

**Remark 7**. *Optimal decision variables don't necessarily depend monotonically on alpha; good decisions are hard to guess and best found by optimization.*

Table 2. Model outputs under varying alpha settings

| case id | case: (demand, disruption) | index | alpha | | |
|---|---|---|---|---|---|
| | | | 10 | 20 | 30 |
| 1 | uniform, exponential-like | BB - # BKP bus | 0 | 0 | 0 |
| | | ITM - z | 0 | 0 | 30 |
| 2 | increasing, normal-like | BB - # BKP bus | 2 | 1 | 0 |
| | | ITM - z | 0 | 0 | 0 |
| 3 | decreasing, bi-Dirac | BB - # BKP bus | 2 | 2 | 0 |
| | | ITM - z | 0 | 10 | 10 |
| 4 | convex, bi-Dirac* | BB - # BKP bus | **2** | **1** | **0** |
| | | ITM - z | **30** | **10** | **10** |
| 5 | increasing, uniform* | BB - # BKP bus | **2** | **1** | **0** |
| | | ITM - z | **10** | **0** | **0** |
| 6 | concave, normal-like* | BM - # BKP bus | **2** | **2** | **1** |
| | | ITM - z | **10** | **30** | **0** |

*\* Seemingly counter-intuitive cases: ITM–z decreases as alpha increases.*

Other studies also focus on high level decisions like network design and resource relocation under disruption. However, those resource relocation methods typically involve more complex frameworks like time-expanded networks (TE-network) and dynamic traffic assignment (DTA) when the costs of users need to be calculated in a precise way under network topology changes caused by disruption, like Jin et al. (2016). The trajectories of individual trains and each user can be generated on a TE-network. These models are expected to achieve better performance on small networks. We argue that our models are simpler but more scalable, meant to be applicable to strategy selection. Compared with using heuristic algorithms for solving previous types of models (like Nikolić and Teodorović (2019)), our models can be solved more efficiently.

## 4 Conclusion

Typical urban transit systems are so complex that any attempt to find the optimal utilization of all resources in a short period of time would encounter great difficulties. We choose to simplify the unit resource from run level to line level so that strategy selection can be optimized at a network level. Two models following this idea are proposed. They differ in the way they handle the uncertainty in disruption duration. When strategies are mapped into resource allocation, the resulting problem is classified as a nonconvex joint routing and resource allocation (nJRRA) problem. We propose a more constrained form that can be solved as a quadratic constrained quadratic programming (QCQP) problem. The assumptions and main ideas of the methodology in this study are summarized below:

- Disruption mitigation decision making is multi-leveled;
- The basic task unit of resource relocation model is average line service level;
- There is a trade-off between the user cost and operator cost;
- Disruption mitigation is a dynamic decision-making process.

To test the models, a quasi-dynamic evaluation program with a given incident duration distribution is constructed using discretized time steps and discrete distributions. FIFO conditions for users are incorporated with dynamic capacity assumptions to determine expected user costs under different strategies. Five different demand patterns and four different disruption distributions are tested on a toy network. The optimal strategies for different combinations of demand pattern and disruption duration distributions are also obtained. Key insights include:

- The overall performance of ITM is the best compared to LLA, BB, and BM among the instances tested, although BM is not far behind and in some cases better.
- When the demand pattern is concave or the duration distribution is bi-Dirac/exponential-like, it is advantageous to postpone the resource relocation decision.
- When users tend to arrive in the near future, or when a disruption is likely to last for a long time, it makes less sense to delay actions.
- When the disruption distribution is exponential-like, a backup bus is not used in BB.
- ITM will not delay for more than an upper limit due to the penalties of user delay.

For future work, system states can be extended to be stochastic and partially observable, and multistage Markov decision processes can be modeled. Overlapping incidents may also be considered, such that resources allocated become unavailable for subsequent disruptions. User responses to mitigation plan could be modeled in a more complex way. User compliance ratios with respect to operator suggestions can be introduced to make the model more realistic. Similar to Pantelidis et al. (2020), one can study multi-operator incentives for working together, but considering post-disruption strategy selection instead of pre-disruption contracts. Other types of mobility operators can also be considered: ridesharing, taxis, etc.

## Acknowledgments

This research was partially funded by the C2SMART University Transportation Center (USDOT #69A3551747124) and NSF CMMI-1652735.

# Appendix: Stochastic disruption duration case test results

Table A1. Performance comparisons under stochastic disruption duration – demand pattern: uniform

| Model | Optimal model objective values | | | Evaluation results | | Comments |
|---|---|---|---|---|---|---|
| | User cost | Operator cost | Total cost | Expected user cost | Total cost | |
| **demand pattern:** | **uniform** | | | **duration distribution:** | | **uniform** |
| LLA | 82417.5 | 0.0 | 82417.5 | 82417.5 | 82417.5 | |
| BB | 80698.8 | 1200.0 | 81898.8 | 80698.8 | 81898.8 | # BU bus = 1 |
| BM | 77317.8 | 2254.9 | 79572.6* | 77317.8 | 79572.6 | |
| ITM | 77660.1 | 2031.3 | 79691.4* | 77641.5 | 79672.8 | **z = 0*** |
| **demand pattern:** | **uniform** | | | **duration distribution:** | | **normal-like** |
| LLA | 82512.0 | 0.0 | 82512.0 | 82539.3 | 82539.3 | |
| BB | 80752.0 | 1200.0 | 81952.0 | 80767.4 | 81967.4 | # BU bus = 1 |
| BM | 77185.3 | 2271.4 | 79456.7* | 77161.1 | 79432.5 | |
| ITM | 77424.2 | 2232.2 | 79656.4* | 77424.2 | 79656.4 | **z = 0*** |
| **demand pattern:** | **uniform** | | | **duration distribution:** | | **exponential-like** |
| LLA | 79551.0 | 0.0 | 79551.0 | 79580.1 | 79580.1 | |
| BB | 79551.0 | 0.0 | 79551.0 | 79580.1 | 79580.1 | # BU bus = 0 |
| BM | 78944.9 | 510.7 | 79455.6* | 78957.6 | 79468.3 | |
| ITM | 79077.2 | 356.0 | 79433.2* | 79215.7 | 79571.7 | **z = 0*** |
| **demand pattern:** | **uniform** | | | **duration distribution:** | | **bi-Dirac** |
| LLA | 82417.5 | 0.0 | 82417.5 | 82417.5 | 82417.5 | |
| BB | 80698.8 | 1200.0 | 81898.8 | 80698.8 | 81898.8 | # BU bus = 1 |
| BM | 77317.8 | 2254.9 | 79572.6* | 77317.8 | **79572.6**** | |
| ITM | 77468.5 | 1241.5 | 78710.0* | 77468.5 | **78710.0**** | z = 10 |

*Run time of 5 minutes reached without convergence;*
*** ITM is significantly better than BM;*
**** Duration distribution is not bi-Dirac, then ITM delay z's are zero.*

Table A2. Performance comparisons under stochastic disruption duration – demand pattern: increasing

| Model | Optimal model objective values | | | Evaluation results | | Comments |
|---|---|---|---|---|---|---|
| | User cost | Operator cost | Total cost | Expected user cost | Total cost | |
| **demand pattern:** | **increasing** | | | **duration distribution:** | | **uniform** |
| LLA | 81674.1 | 0.0 | 81674.1 | 81911.6 | 81911.6 | |
| BB | 79955.4 | 1200.0 | 81155.4 | 80243.5 | 81443.5 | # BU bus = 1 |
| BM | 78641.9 | 2235.8 | 80877.7* | 78659.0 | 80894.9 | |
| ITM | 77297.1 | 2228.8 | 79525.9* | 77297.2 | 79525.9 | z = 0 |
| **demand pattern:** | **increasing** | | | **duration distribution:** | | **normal-like** |
| LLA | 81770.2 | 0.0 | 81770.2 | 81934.3 | 81934.3 | |
| BB | 80010.2 | 1200.0 | 81210.2 | 80230.1 | 81430.1 | # BU bus = 1 |
| BM | 77494.5 | 2010.1 | 79504.7* | 77455.9 | 79466.1 | |
| ITM | 77213.5 | 2378.2 | 79591.6* | 77213.5 | 79591.6 | z = 0 |
| **demand pattern:** | **increasing** | | | **duration distribution:** | | **exponential-like** |
| LLA | 79134.1 | 0.0 | 79134.1 | 79193.9 | 79193.9 | |
| BB | 79134.1 | 0.0 | 79134.1 | 79193.9 | 79193.9 | # BU bus = 0 |
| BM | 79134.1 | 0.0 | 79134.1* | 79193.9 | 79193.9 | |
| ITM | 78977.3 | 194.3 | 79171.5* | 78988.1 | 79182.4 | **z = 20**** |
| **demand pattern:** | **increasing** | | | **duration distribution:** | | **bi-Dirac** |
| LLA | 81674.1 | 0.0 | 81674.1 | 82358.1 | 82358.1 | |
| BB | 79955.4 | 1200.0 | 81155.4 | 80588.6 | 81788.6 | # BU bus = 1 |
| BM | 78641.9 | 2235.8 | 80877.7* | 78689.3 | 80925.1 | |
| ITM | 77428.1 | 1247.8 | 78675.9* | 77428.1 | 78675.9 | **z = 20**** |

\* *Run time of 5 minutes reached without convergence;*
\*\* *When the demand pattern is increasing, it is advantageous to postpone the resource relocation decision.*

Table A3. Performance comparisons under stochastic disruption duration – demand pattern: decreasing

| Model | Optimal model objective values | | | Evaluation results | | Comments |
|---|---|---|---|---|---|---|
| | User cost | Operator cost | Total cost | Expected user cost | Total cost | |
| **demand pattern:** | **decreasing** | | | **duration distribution:** | | **uniform** |
| LLA | 83310.9 | 0.0 | 83310.9 | 83147.4 | 83147.4 | |
| BB | 79895.3 | 2400.0 | 82295.3 | 79835.4 | 82235.4 | # BU bus = 2 |
| BM | 77063.7 | 2341.4 | 79405.1* | 77133.6 | 79475.0 | |
| ITM | 77320.2 | 2187.2 | 79507.4* | 77320.2 | 79507.4 | z = 0 |
| **demand pattern:** | **decreasing** | | | **duration distribution:** | | **normal-like** |
| LLA | 83394.6 | 0.0 | 83394.6 | 83367.4 | 83367.4 | |
| BB | 79921.8 | 2400.0 | 82321.8 | 79894.7 | 82294.7 | **# BU bus = 2**** |
| BM | 77271.2 | 2790.5 | 80061.7* | 77410.9 | 80201.4 | |
| ITM | 76967.9 | 2583.0 | 79550.9* | 76968.0 | 79551.0 | z = 0 |
| **demand pattern:** | **decreasing** | | | **duration distribution:** | | **exponential-like** |
| LLA | 80111.8 | 0.0 | 80111.8 | 80089.0 | 80089.0 | |
| BB | 80111.8 | 0.0 | 80111.8 | 80089.0 | 80089.0 | # BU bus = 0 |
| BM | 79102.4 | 622.3 | 79724.7* | 79109.8 | 79732.1 | |
| ITM | 79417.0 | 326.6 | 79743.6* | 79666.0 | 79992.6 | z = 0 |
| **demand pattern:** | **decreasing** | | | **duration distribution:** | | **bi-Dirac** |
| LLA | 83310.9 | 0.0 | 83310.9 | 82715.4 | 82715.4 | |
| BB | 79895.3 | 2400.0 | 82295.3 | 79729.9 | 82129.9 | **# BU bus = 2**** |
| BM | 77063.7 | 2341.4 | 79405.1* | 77256.7 | 79598.1 | |
| ITM | 77937.8 | 1252.7 | 79190.5* | 77937.8 | 79190.5 | z = 10 |

\* *Run time of 5 minutes reached without convergence*
\*\* *Backup buses are heavily used when the demand pattern is concave.*

Table A4. Performance comparisons under stochastic disruption duration – demand pattern: convex

| Model | Optimal model objective values | | | Evaluation results | | Comments |
|---|---|---|---|---|---|---|
| | User cost | Operator cost | Total cost | Expected user cost | Total cost | |
| **demand pattern:** | **convex** | | | **duration distribution:** | | **uniform** |
| LLA | 73225.1 | 0.0 | 73225.1 | 73318.6 | 73318.6 | |
| BB | 71506.3 | 1200.0 | 72706.3 | 71655.8 | 72855.8 | # BU bus = 1 |
| BM | 68824.2 | 2114.5 | 70938.8* | 68814.8 | 70929.3 | |
| ITM | 68684.5 | 2255.5 | 70940.0* | 68684.5 | 70940.0 | z = 0 |
| **demand pattern:** | **convex** | | | **duration distribution:** | | **normal-like** |
| LLA | 73288.2 | 0.0 | 73288.2 | 73396.4 | 73396.4 | |
| BB | 71528.2 | 1200.0 | 72728.2 | 71690.1 | 72890.1 | # BU bus = 1 |
| BM | 68781.1 | 2087.2 | 70868.3* | 68766.9 | 70854.1 | |
| ITM | 68902.4 | 2015.3 | 70917.7* | 68902.4 | 70917.7 | z = 0 |
| **demand pattern:** | **convex** | | | **duration distribution:** | | **exponential-like** |
| LLA | 71046.6 | 0.0 | 71046.6 | 71003.0 | 71003.0 | |
| BB | 71046.6 | 0.0 | 71046.6 | 71003.0 | 71003.0 | # BU bus = 0 |
| BM | 70248.5 | 571.7 | 70820.2* | 70253.2 | 70824.9 | |
| ITM | 70367.5 | 384.7 | 70752.1* | 70533.0 | 70917.6 | z = 0 |
| **demand pattern:** | **convex** | | | **duration distribution:** | | **bi-Dirac** |
| LLA | 73225.1 | 0.0 | 73225.1 | 73398.0 | 73398.0 | |
| BB | 71506.3 | 1200.0 | 72706.3 | 71712.7 | 72912.7 | # BU bus = 1 |
| BM | 68824.2 | 2114.5 | 70938.8* | 68798.2 | 70912.7 | |
| ITM | 69231.1 | 1150.2 | 70381.3* | 69231.1 | 70381.3 | z = 10 |

*Run time of 5 minutes reached without convergence*

Table A5. Performance comparisons under stochastic disruption duration – demand pattern: concave

| Model | Optimal model objective values | | | Evaluation results | | Comments |
|---|---|---|---|---|---|---|
| | User cost | Operator cost | Total cost | Expected user cost | Total cost | |
| **demand pattern:** | **concave** | | | **duration distribution:** | | **uniform** |
| LLA | 91636.6 | 0.0 | 91636.6 | 91831.6 | 91831.6 | |
| BB | 88460.0 | 2400.0 | 90860.0 | 88450.0 | 90850.0 | **# BU bus = 2\*\*\*** |
| BM | 85651.1 | 2361.2 | 88012.3* | 85663.4 | 88024.5 | |
| ITM | 85704.6 | 2327.4 | 88032.0* | 85704.6 | 88032.0 | z = 0 |
| **demand pattern:** | **concave** | | | **duration distribution:** | | **normal-like** |
| LLA | 91778.2 | 0.0 | 91778.2 | 92013.7 | 92013.7 | |
| BB | 88495.9 | 2400.0 | 90895.9 | 88497.0 | 90897.0 | # BU bus = 2*** |
| BM | 85631.2 | 2344.0 | 87975.2* | 85629.9 | 87974.0 | |
| ITM | 86421.6 | 2225.5 | 88647.1* | 86420.8 | 88646.2 | z = 30 |

| | demand pattern: | concave | | duration distribution: | | exponential-like |
|---|---|---|---|---|---|---|
| LLA | 88097.2 | 0.0 | 88097.2 | 88226.1 | 88226.1 | |
| BB | 88097.2 | 0.0 | 88097.2 | 88226.1 | 88226.1 | # BU bus = 0 |
| BM | 87642.6 | 402.1 | 88044.7* | 87766.2 | 88168.4 | |
| ITM | 87833.6 | 269.8 | 88103.4* | 87919.9 | 88189.7 | **z = 10**** |
| | **demand pattern:** | **concave** | | **duration distribution:** | | **bi-Dirac** |
| LLA | 91636.6 | 0.0 | 91636.6 | 91759.8 | 91759.8 | |
| BB | 88460.0 | 2400.0 | 90860.0 | 88432.4 | 90832.4 | **# BU bus = 2**** |
| BM | 85651.1 | 2361.2 | 88012.3* | 85685.0 | 88046.1 | |
| ITM | 86008.0 | 1206.0 | 87214.0* | 86008.0 | 87214.0 | **z = 10**** |

*Run time of 5 minutes reached without convergence;*
*** When duration distribution is bi-Dirac/exponential-like, it is advantageous to postpone the resource relocation decision.*
**** Backup buses are heavily used when the demand pattern is concave.*